\documentclass[twocolumn,amsmath,amssymb,aps,groupedaddress,nofootinbib,pra]{revtex4-2}
\usepackage{amsthm}
\usepackage{graphicx}
\usepackage{bbm}
\usepackage{array}
\newcolumntype{C}{>{$}c<{$}}
\AtBeginDocument{
\heavyrulewidth=.08em
\lightrulewidth=.05em
\cmidrulewidth=.03em
\belowrulesep=.65ex
\belowbottomsep=0pt
\aboverulesep=.4ex
\abovetopsep=0pt
\cmidrulesep=\doublerulesep
\cmidrulekern=.5em
\defaultaddspace=.5em
\tabcolsep=7pt
}

\usepackage[hidelinks]{hyperref}
\usepackage{booktabs}
\usepackage{comment}
\usepackage{siunitx}
\usepackage[braket, qm]{qcircuit}
\usepackage{natbib}

\usepackage{makecell, multirow, tabularx}
\begin{document}

\title{Fault-Tolerant Implementation of the Deutsch-Jozsa Algorithm}

\author{Divyanshu Singh}
\email{divyanshusingh239@gmail.com}
\author{Shiroman Prakash}
\email{sprakash@dei.ac.in}
\affiliation{Department of Physics and Computer Science, Dayalbagh Educational Institute, Agra, India}
\begin{abstract}
 We show that one can implement the Deutsch-Josza algorithm, one of the first and simplest quantum algorithms, in a fault-tolerant manner using the smallest quantum error-detecting code -- the $[[4,2,2]]$ code -- without any ancillae. We implemented the algorithm on a trapped-ion quantum computer with and without fault-tolerant encoding and compared the results. With approximately $99 \%$ confidence, we found that the fault-tolerant implementation provided a noise reduction for all oracles. Averaged across all oracles, the reduction in error rate was nearly $90 \%$. 

\end{abstract}

\maketitle
\section{Introduction}

Quantum computation holds immense promise (e.g., \cite{Shor:1994jg, grover1996fast, DPS2011,DPS2014, DPS2, DPS1, quantum-algorithms}), but realizing its full technological potential requires fault-tolerance \cite{Shor1996qc, Gottesman:1997qd}. Several long-term strategies for achieving large-scale fault-tolerance exist, perhaps most notably, using the surface code or Majorana fermions, supplemented with magic state distillation, but successful realization of these techniques require substantial advances in hardware, and it appears unlikely that fault-tolerant techniques can be useful in presently-available quantum devices. However, as experimentally-realized noise rates reduce, successful demonstrations of simple fault-tolerant protocols \cite{gottesman2016quantum} are entering the realm of possibility (e.g., \cite{Vuillot2018, HarperFlammia2019, Pokharel:2022vwm, Wang:2023qcn, Menendez:2023veg, Postler:2023jex, reichardt2024demonstration}). It would be satisfying to use fault-tolerant techniques to solve a well-defined problem on a quantum computer with an unambiguously lower error-rate than one can achieve without fault-tolerance. The following natural question therefore arises: Is it possible to implement a textbook quantum algorithm in a completely fault-tolerant manner? If so, is the noise reduction substantial? 

Here, we give an affirmative answer to both these questions. We provide a completely fault-tolerant implementation of the first quantum algorithm -- the Deutsch-Josza algorithm \cite{Deutsch1985, deutsch-jozsa} -- using the smallest qubit error-detecting code -- the $[[4,2,2]]$ code \cite{Vaidman:1996bg} (see also \cite{Grassl:1996eh}), and demonstrate that it provides a statistically significant $90 \%$ reduction in noise for all possible inputs on an ion-trap quantum computer, IonQ's Aria-1 \cite{Aria-1}. We emphasize at the outset that our implementation is extremely simple -- so simple, some may argue it is trivial -- however, given the fact that the present work appears to be the first (and simplest possible) completely fault-tolerant implementation of any quantum algorithm, we believe it may be of some value to the community. 

\begin{figure}[h]
    \centering
    \scalebox{1.0}{
    \Qcircuit @C=1.0em @R=0.2em @!R { \\
    	 	\nghost{{q}_{1} :  } & \lstick{{q}_{1} :  } & \gate{X} & \gate{H} \barrier[0em]{1} & \qw & \multigate{1}{\mathrm{oracle}}_<<<{1} \barrier[0em]{1} & \qw & \qw  & \qw \\
    	 	\nghost{{q}_{2} :  } & \lstick{{q}_{2} :  } & \gate{X} & \gate{H} & \qw & \ghost{\mathrm{oracle}}_<<<{2} & \qw &  \gate{H}  & \meter \\
    \\ }}
    \caption{A circuit implementing the Deutsch-Josza algorithm. With a single query, the output of the measurement on the last qubit indicates whether the function encoded by the oracle is balanced or constant.}
    \label{fig:bare-deutsch-jozsa}
\end{figure}
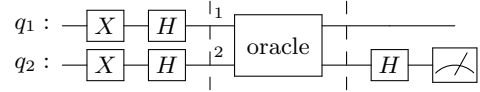

The Deutsch-Josza algorithm is often the first quantum algorithm introduced in standard textbooks (e.g.,\cite{Mermin_2007, Nielsen_Chuang_2010}). It is an oracle-based quantum algorithm, where the computational advantage arises from reducing the number of oracle queries required to solve a problem posed by Deutsch \cite{Deutsch1985}. There are two parties, Alice and Bob. Alice prepares an oracle implementing a binary function on a single bit $f(x)$. Bob needs to ascertain whether the function $f(x)$ is constant or balanced using as few oracle queries as possible. Classically, Bob requires two oracle queries to solve this problem, however, using quantum superposition, Deutsch and Josza \cite{deutsch-jozsa} showed it is possible to determine whether $f(x)$ is constant or balanced using a single oracle query. The circuit implementing the Deutsch-Josza algorithm is shown in Figure \ref{fig:bare-deutsch-jozsa}. See \cite{Cleve:1997dh}
for more discussion. 

The circuit implementing the Deutsch-Josza algorithm, as well as the oracles implementing boolean functions $f(x)$ of a single-bit $x$,  utilize only Clifford gates for their implementation. The Deutsch-Josza algorithm is therefore a natural candidate for fault-tolerant implementation using currently available quantum computers.\footnote{Recall that Clifford unitaries are those unitaries which map Pauli operators to Pauli operators under conjugation, and are, therefore, by definition, easier to implement as transversal gates in stabilizer codes than non-Clifford gates. The Deutsch-Josza algorithm can be generalized to $n$-bit binary functions; and, in general, the oracles require non-Clifford gates for their implementation.}

 The $[[4,2,2]]$ code of \cite{Vaidman:1996bg}, being the smallest quantum code in existence, is the naturally the most attractive candidate for early experiments in fault-tolerance \cite{gottesman2016quantum}. Of course, has a limited number of transversal gates, so one can ask, can one do anything interesting with this code? Several groups have utilized this code for experiments that could be classified as benchmarking, including Vuillot \cite{Vuillot2018} and Willsch \textit{et al.} \cite{willsch2018testing} -- who studied and attempted (with mixed results) to demonstrate fault tolerance in early superconducting devices of IBM; Linke \textit{et al} \cite{ion_trap_fault_tolerance} and Takita \textit{et al} \cite{Takita:2017blo} -- who demonstrated fault-tolerant error-detection and state preparation using the $[[4,2,2]]$ code using ion trap and superconducting qubits, respectively; Harper and Flammia \cite{HarperFlammia2019} -- who used the $[[4,2,2]]$ code to demonstrate a noise reduction in execution of gates, but not state preparation and measurement. More recently, Pokharel and Lidar  \cite{Pokharel:2022vwm} used the $[[4,2,2]]$ code to reduce noise in an implementation of Grover's algorithm that was not, however, completely fault-tolerant, and largely relied on error mitigation. Our work is inspired from those mentioned above, but differs in that we use the $[[4,2,2]]$ code to provide a completely fault-tolerant implementation of a textbook quantum algorithm and demonstrate a statistically significant reduction in noise for all possible inputs. 

 For context, we should also mention \cite{Wang:2023qcn} which implements the non-trivial primitive operation one-qubit addition fault-tolerantly using the $[[8,1,2]]$ colour code, \cite{Menendez:2023veg} which implements non-Clifford gates using the $[[8,1,2]]$ colour code. 

Let us remark that, in this work and those cited above, noise-suppression arises via error detection rather than error-correction. The utility of error-detection in fault-tolerance is sometimes controversial, but can be made transparent in this setting, if Deutsch's problem can be viewed as a game with three possible outcomes for Bob -- a win, loss or draw. Bob is given only one oracle query. He wins or loses depending on whether he correctly or incorrectly ascertains whether Alice's secret binary function is balanced or constant, and draws if chooses to abstain from giving an answer. If Bob detects that an error has occurred in his circuit, he knows that its output cannot be trusted, so he is able reduce his loss-rate by abstaining. If Bob does not have the third option of abstaining, error-detection provides him no advantage, and error-correction is required to increase his win-rate. Experiments with error-correction are somewhat challenging, but we should mention \cite{Postler:2023jex} who carry out successful fault-tolerant error correction using the $[[7,1,3]]$ Steane code. 

We chose to implement our circuits on IonQ's 25-qubit ion-trap quantum computer \cite{iontrap}, Aria-1. However, due to the extreme simplicity of our protocol, we expect that similar reductions in noise using the circuits provided in this paper could be obtained using a variety of quantum computers currently available, including superconducting qubits.

We should emphasize that the quantity we measure -- the amount of noise reduction from a fault-tolerant implementation -- is extremely difficult to calculate theoretically, unless one makes numerous simplifying assumptions. 
%In any fault-tolerant implementation, an important consideration is the number of ancillae required. Somewhat surprisingly, we find that it is possible to implement the Deutsch-Josza algorithm in a completely-fault-tolerant manner with no ancillae. 

\section{Methods}
%We now present a fault-tolerant implementation of the Deutsch-Josza algorithm in the $[[4,2,2]]$ error-detecting code that requires no ancillae. 

\subsection{General strategy}
We will compare a bare (non-fault tolerant) implementation of the Deutsch-Josza algorithm to a fault-tolerant implementation, using the $[[4,2,2]]$ code. The bare implementation utilizes only two qubits and implements the bare circuits in Figure \ref{fig:bare-deutsch-jozsa}, and bare oracles in the second column of Table \ref{tab:fault-tolerant-oracles}, without making use of any error-detecting or correcting code. These circuits are described in many textbooks, e.g., \cite{Mermin_2007}.

What constitutes a fault-tolerant implementation of the Deutsch-Josza algorithm? Alice supplies Bob with an oracle implementing one of the four binary functions. We assume Alice agrees to provide a fault-tolerant version of this oracle, using a specified dictionary, which in our case will be the dictionary in Table \ref{tab:transversal-gates}. Bob will then apply a fault-tolerant encoding of the circuit implementing the Deutsch-Josza algorithm to the oracle. Based on his measured result, he will decide whether or not the function is balanced or constant. Crucially, Bob does not know what oracle Alice has chosen, so his circuit must be fault-tolerant regardless of which of the four oracles Alice chooses to give him. We also allow Alice and Bob to individually simplify their respective quantum circuits. However, the oracles and Deutsch-Josza circuit must be prepared and executed independently, so we demand that no simplification occurs that combines gates from both Alice's and Bob's circuits. 

Because of the limited number of transversal gates associated with the $[[4,2,2]]$ code, it is not obvious one can implement the Deutsch-Josza algorithm using it. We show that a fault-tolerant implementation of the Deutsch-Josza algorithm is possible with no ancillae, using the particular presentation of the $[[4,2,2]]$ code in the next subsection.

\subsection{$[[4,2,2]]$ code}

The $[[4,2,2]]$ code \cite{Vaidman:1996bg, Grassl:1996eh}  is defined by the stabilizers $S_1=XXXX$ and $S_2=ZZZZ$. We choose logical operators $\overline{X}_1 =ZZII$, $\overline{Z}_1=XIXI$, $\overline{X}_2=XXII$ and $\overline{Z}_2=ZIZI$. This choice, which differs slightly from the choice made by other authors \cite{gottesman2016quantum,  Vuillot2018, HarperFlammia2019, Pokharel:2022vwm}, enables one to fault-tolerantly implement the Deutsch-Josza algorithm without any ancillae.\footnote{If we were to directly use the logical operators chosen in \cite{gottesman2016quantum} to implement the Deutsch-Josza algorithm fault-tolerantly, one would require 1 ancilla as well as 5 CNOT gates.} Table \ref{tab:transversal-gates} presents some transversal gates for the $[[4,2,2]]$ code, with this choice of logical gates. 

The logical states for this code are
\begin{equation}
\begin{split}
\overline{\ket{+0}}  & =  \frac{1}{\sqrt{2}} \left( \ket{0000} + \ket{1111} \right),\\
\overline{\ket{+1}}  & =  \frac{1}{\sqrt{2}} \left( \ket{1100} + \ket{0011} \right), \\
\overline{\ket{-0}}  & =  \frac{1}{\sqrt{2}} \left( \ket{1010} + \ket{0101}  \right),  \\
\overline{\ket{-1}}  & =  \frac{1}{\sqrt{2}} \left( \ket{0110} + \ket{1001}\right). \label{logical-states}
\end{split}
\end{equation}

As in \cite{gottesman2016quantum}, we can implement a fault-tolerant simultaneous measurement of $\overline{X}_1$ and $\overline{Z}_2$ without any ancillae by measuring each of the four physical qubits in the computational basis. Measurement outcomes for the physical qubits are translated into outcomes for the logical qubits via equation \eqref{logical-states}. A single $X$ error on any physical qubit will cause the total number of $1$'s measured to be odd, in which case we discard the run. A single $Z$ error on any physical qubit, while undetectable, will not affect the measurement results and can be safely ignored.

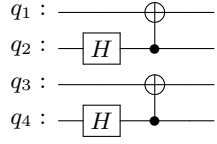
\begin{figure}
    \centering
    \scalebox{1.0}{
    \Qcircuit @C=1.0em @R=0.2em @!R { \\
    	 	\nghost{{q}_{1} :  } & \lstick{{q}_{1} :  } & \qw & \targ & \qw & \qw\\
    	 	\nghost{{q}_{2} :  } & \lstick{{q}_{2} :  } & \gate{H} & \ctrl{-1} & \qw & \qw\\
    	 	\nghost{{q}_{3} :  } & \lstick{{q}_{3} :  } & \qw & \targ & \qw & \qw\\
    	 	\nghost{{q}_{4} :  } & \lstick{{q}_{4} :  } & \gate{H} & \ctrl{-1} & \qw & \qw\\
    \\ }}
    \caption{Fault tolerant preparation of the state $\overline{\ket{++}}$ in the $[[4,2,2]]$ error-detecting code. }
    \label{fig:state-preparation}
\end{figure}

We can also fault-tolerantly prepare the logical state $\overline{\ket{++}}$, which can be expressed as the product of Bell states, \begin{equation} \begin{split}
   \overline{\ket{++}} & = \frac{1}{2} \left( \ket{0000} + \ket{0011}+\ket{1100}+\ket{1111}\right) \\
    & =\frac{1}{2} \left( \ket{00} + \ket{11}\right)\left(\ket{00}+\ket{11}\right), \label{initial-state}
   \end{split}
\end{equation} using the circuit in Figure \ref{fig:state-preparation}. As observed in \cite{gottesman2016quantum}, to see that this circuit is fault-tolerant, note that an error in one of the CNOT gates could induce a two-qubit error on one of the Bell states, which can be written as a linear superposition of Pauli errors. However, using $U_1\otimes U_2 \left(\ket{00} + \ket{11}\right)= 1\otimes U_2U_1^T \left( \ket{00} + \ket{11}\right)$, any two-qubit Pauli error on the Bell state is equivalent to a single qubit error. Of course, to prepare other logical states, such as $\ket{+}\ket{0}$, we would in general require an ancilla.

\begin{table}
\small
\renewcommand{\arraystretch}{1.2}
\begin{tabular}{c c c c c}
\toprule
\textbf{Logical Gate}&\textbf{Physical Gate(s)}\\ \midrule
$\overline{I}\otimes \overline{I}$&$X^{\otimes 4}, Z^{\otimes 4}$ \\
$\overline{X} \otimes \bar I$&$Z \otimes Z \otimes I \otimes I$ \\
$\overline{Z}\otimes \overline{I}$&$X \otimes I \otimes X \otimes I$ \\
$\bar I\otimes \overline{X}$&$X \otimes X \otimes I \otimes I$  \\
$\bar I\otimes \overline{Z}$&$Z \otimes I \otimes Z \otimes I$  \\
$\overline{\rm CNOT}_{21}$ &$S \otimes ZS \otimes ZS \otimes S$ \\
$\overline{\rm SWAP}_{12}$ &$H \otimes H \otimes H \otimes H$  \\
 %\midrule
%$(H\otimes I)\circ$(CNOT$_{12}$)$\circ(H\otimes I)$&SWAP$_{12}$  \\
%C-Z&SWAP$_{13}$  \\
%SWAP$_{12} \circ (H\otimes H)$&SWAP$_{23}$  \\
\bottomrule
\end{tabular}
\caption{A useful choice of logical operators for the [[4,2,2]] code, with corresponding transversal gates.} \label{tab:transversal-gates}
%\end{center}
\end{table}

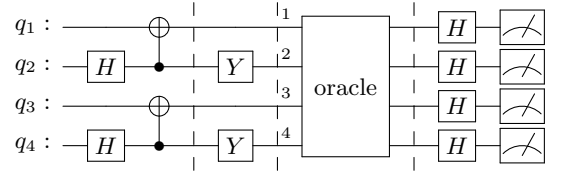
\begin{figure}
    \centering
    \scalebox{1.0}{
    \Qcircuit @C=1.0em @R=0.2em @!R { \\
    	 	\nghost{{q}_{1} :  } & \lstick{{q}_{1} :  } & \qw & \targ \barrier[0em]{3} & \qw & \qw \barrier[0em]{3} & \qw & \multigate{3}{\mathrm{oracle}}_<<<{1} \barrier[0em]{3} & \qw & \gate{H} & \meter\\
    	 	\nghost{{q}_{2} :  } & \lstick{{q}_{2} :  } & \gate{H} & \ctrl{-1} & \qw & \gate{Y} & \qw & \ghost{\mathrm{oracle}}_<<<{2} & \qw & \gate{H} &  \meter \\
    	 	\nghost{{q}_{3} :  } & \lstick{{q}_{3} :  } & \qw & \targ & \qw & \qw & \qw & \ghost{\mathrm{oracle}}_<<<{3} & \qw & \gate{H} &  \meter \\
    	 	\nghost{{q}_{4} :  } & \lstick{{q}_{4} :  } & \gate{H} & \ctrl{-1} & \qw & \gate{Y} & \qw & \ghost{\mathrm{oracle}}_<<<{4} & \qw & \gate{H} &  \meter \\
    \\ }}
    \caption{Fault-tolerant encoding of the Deutsch-Jozsa algorithm in the $[[4,2,2]]$ code. In the last step, we swap the logical qubits before measurement, in order to measure $X$ rather than $Z$ on the logical qubit of interest.}
    \label{fig:encoded-deutsch-jozsa}
\end{figure}

A fault-tolerant implementation of the Deutsch-Josza algorithm within the $[[4,2,2]]$ code, without any ancillae, is shown in Figure \ref{fig:encoded-deutsch-jozsa}, which crucially takes advantage of the ability to fault-tolerantly prepare the state $\overline{\ket{++}}$ and measure one of the qubits in the $X$ eigenbasis. Fault-tolerant implementations of oracles corresponding to each of the four possible binary functions $f(x)$ using transversal gates are shown in Table \ref{tab:fault-tolerant-oracles}.  The resulting circuits are fault-tolerant -- if a single error occurs on any qubit, gate, or measurement that would affect the final results, it will be detected and the run discarded.

\begin{table}[h]
    \centering
    \begin{tabular}{  m{2cm}  m{8em} m{9em}  } 
    \toprule
    \textbf{\makecell{Binary \\ Function}} & \textbf{\makecell{Bare \\ Oracle}} & \textbf{\makecell{Encoded \\ Oracle}} \\ \midrule
    \multirow{2.7}{*}{\hspace{1em}$f(x) = 0$} & \vspace{1em}
    \scalebox{1.0}{
    \Qcircuit @C=0.6em @R=0.2em @!R {
        \nghost{{q}_{1} :  } & \lstick{{q}_{1} :  } & \qw & \qw\\
        \nghost{{q}_{2} :  } & \lstick{{q}_{2} :  } & \qw & \qw\\
    }} & 
    \scalebox{1.2}{
    \Qcircuit @C=0.6em @R=0.2em @!R {
        \nghost{{q}_{1} :  } & \lstick{{q}_{1} :  } & \qw & \qw\\
        \nghost{{q}_{2} :  } & \lstick{{q}_{2} :  } & \qw & \qw\\
        \nghost{{q}_{3} :  } & \lstick{{q}_{3} :  } & \qw & \qw\\
        \nghost{{q}_{4} :  } & \lstick{{q}_{4} :  } & \qw & \qw\\
    }} \\ 
    \multirow{2.7}{*}{\hspace{1em}$f(x) = x$} & \vspace{1em}
    \scalebox{1.0}{
    \Qcircuit @C=0.6em @R=0.2em @!R {
        \nghost{{q}_{1} :  } & \lstick{{q}_{1} :  } & \targ & \qw & \qw\\
        \nghost{{q}_{2} :  } & \lstick{{q}_{2} :  } & \ctrl{-1} & \qw & \qw\\
    }} &
    \scalebox{1.0}{
    \Qcircuit @C=0.6em @R=0.2em @!R {
        \nghost{{q}_{1} :  } & \lstick{{q}_{1} :  } & \gate{S} & \qw & \qw & \qw\\
        \nghost{{q}_{2} :  } & \lstick{{q}_{2} :  } & \gate{Z} & \gate{S} & \qw & \qw\\
        \nghost{{q}_{3} :  } & \lstick{{q}_{3} :  } & \gate{Z} & \gate{S} & \qw & \qw\\
        \nghost{{q}_{4} :  } & \lstick{{q}_{4} :  } & \gate{S} & \qw & \qw & \qw\\
    }} \\ 
    \multirow{2.7}{*}{\hspace{0.2em}$f(x) = 1 \oplus x$}& \vspace{1em}
    \scalebox{1.0}{
    \Qcircuit @C=0.6em @R=0.2em @!R {
        \nghost{{q}_{1} :  } & \lstick{{q}_{1} :  } & \gate{X} & \targ & \qw & \qw\\
        \nghost{{q}_{2} :  } & \lstick{{q}_{2} :  } & \qw & \ctrl{-1} & \qw & \qw\\
    }} &
    \scalebox{1.0}{
    \Qcircuit @C=0.6em @R=0.1em @!R {
        \nghost{{q}_{1} :  } & \lstick{{q}_{1} :  } & \gate{Z} & \gate{S} & \qw & \qw\\
        \nghost{{q}_{2} :  } & \lstick{{q}_{2} :  } & \gate{S} & \qw & \qw & \qw\\
        \nghost{{q}_{3} :  } & \lstick{{q}_{3} :  } & \gate{Z} & \gate{S} & \qw & \qw\\
        \nghost{{q}_{4} :  } & \lstick{{q}_{4} :  } & \gate{S} & \qw & \qw & \qw\\
    }} \\ 
    \multirow{2.7}{*}{\hspace{1em}$f(x) = 1$} & \vspace{1em}
    \scalebox{1.0}{
    \Qcircuit @C=0.6em @R=0.2em @!R {
        \nghost{{q}_{1} :  } & \lstick{{q}_{1} :  } & \gate{X} & \qw & \qw\\
        \nghost{{q}_{2} :  } & \lstick{{q}_{2} :  } & \qw & \qw & \qw\\ 
    }} &
    \scalebox{1.0}{
    \Qcircuit @C=0.6em @R=0.2em @!R {
        \nghost{{q}_{1} :  } & \lstick{{q}_{1} :  } & \gate{Z} & \qw & \qw\\
        \nghost{{q}_{2} :  } & \lstick{{q}_{2} :  } & \gate{Z} & \qw & \qw\\
        \nghost{{q}_{3} :  } & \lstick{{q}_{3} :  } & \qw & \qw & \qw\\
        \nghost{{q}_{4} :  } & \lstick{{q}_{4} :  } & \qw & \qw & \qw\\
    }} \\ \bottomrule
    \end{tabular}
    \caption{Bare and encoded oracle implementations.}
    \label{tab:fault-tolerant-oracles}
\end{table}

\subsection{Statistical Distance}
Let $P_i$ denote the theoretical probability of outcome labeled $i$ in an ideal quantum circuit, $Q_i$ denote the observed probability of the outcome $i$ in the bare (non-fault-tolerant) quantum circuit, and $R_i$ the observed probability in the encoded (fault-tolerant) circuit. 

In the bare circuit implementing the Deutsch-Josza algorithm, Figure \ref{fig:bare-deutsch-jozsa}, the output is obtained by measuring the second qubit only. If the oracle implements a constant function then $P_0=0$ and $P_1=1$, and if the oracle implements a balanced function then $P_0=1$ and $P_1=0$

When we run the algorithm, we measure the second qubit and obtain probabilities $Q_0$ and $Q_1$,
which may be compared to the ideal probabilities $P_0$ and $P_1$, via the statistical distance, defined in \cite{gottesman2016quantum} to be,
\begin{align}
  D_{\rm bare}=\frac{1}{2}(|P_{0}-Q_{0}|+|P_{1}-Q_{1}|). 
\end{align}

When we run the encoded circuits, we measure all four qubits, resulting in probabilities $R_{ijkl}$ for 16 possible outcomes. The error detection step consists of discarding those outcomes outside of the codespace (\textit{i.e.}, those containing an odd number of $1$'s), to obtain $8$ probabilities $R_{ijkl}$, which are re-normalized to sum to $1$. These are then translated to probabilities for logical qubits measurement via,
\begin{equation}
\begin{split}
    {R_{00}} & ={R'_{0000}}+{R'_{1111}},  \\
    {R_{01}}& ={R'_{1100}}+{R'_{0011}}, \\
    {R_{10}}& ={R'_{1010}}+{R'_{0101}}, \\
    {R_{11}}& ={R'_{0110}}+{R'_{1001}}.
    \end{split}
\end{equation}
In the encoded circuit, Figure \ref{fig:encoded-deutsch-jozsa}, it was necessary to swap the logical qubits prior to measurement. We then have $R_0=R_{00}+R_{01}$ and $R_{1}=R_{10}+R_{11}$, and the statistical distance for the encoded circuit is
\begin{align}
    D_{\rm enc}=\frac{1}{2}(|P_{0}-R_{0}|+|P_{1}-R_{1}|).
\end{align}
The fraction of runs that were not discarded is the post-selection ratio.

\section{Results}
We ran the Deutsch-Josza algorithm on all four oracles with and without the fault-tolerant encoding described above. We implemented fault-tolerant circuits using native gates \cite{native-gates}, and disabled error mitigation and circuit post-processing using the options available in Amazon braket \cite{verbatim-compilation}, as described in Appendix \ref{app:nativegate}. Each run consisted of 4096 shots.

\begin{table*}[ht]
	\renewcommand{\arraystretch}{2}
	\centering
	\vspace{0.1cm}
	\begin{tabular}{l  c  c  c  c }
		\toprule
		 \textbf{Oracle}& \textbf{\makecell{Statistical Distance \\ (Bare)}} & \textbf{\makecell{Statistical Distance\\ (Encoded)}}& \textbf{\makecell{Statistical Distance\\ (Encoded-Bare)}} & \textbf{\makecell{Post-Selection \\Ratio}}\\
            \midrule 
		$f(x)=0$& $0.01929 \pm 0.00152$ & $0.00178 \pm 0.00048$ & $-0.01750  \pm 0.00159$ &0.958\\

		$f(x)=x$ &$0.00317 \pm 0.00062$&$0.00128 \pm 0.00040$&	$-0.00189 \pm 0.00074$&0.954\\
	   $f(x)=1\oplus x$ &$0.00513 \pm 0.00078$&	$0.00076 \pm 0.00031$	&$-0.00436 \pm 0.00085$&0.958\\

		$f(x)=1$ &$0.01782 \pm 0.00146$&$0.00129 \pm 0.00040$&$-0.01654 \pm 0.00152$&0.950\\
	\bottomrule 
	\end{tabular}
    \caption{Results for bare (non-fault tolerant) and encoded (fault-tolerant) implementation of the Deutsch-Josza algorithm for all oracles, on IonQ's Aria-1 machine, with 4096 shots in each run. }
    \label{tab:results}
\end{table*}

In order to compare the results between bare and encoded circuits, \cite{gottesman2016quantum} defined the  \textit{statistical distance}, which compares observed probability distribution of outcomes for a given circuit with the ideal probability distribution expected for that circuit in the absence of noise, as a performance metric. In our opinion, the statistical distance is not necessarily the most meaningful metric from the perspective of the benchmarking experiments of \cite{gottesman2016quantum, Vuillot2018, HarperFlammia2019}, and a complete characterization of the noise reduction associated with a fault-tolerant scheme should ideally include detailed tomography studies, for the purpose of evaluating a fault-tolerant implementation of a particular quantum algorithm, the statistical distance is the most natural performance metric to use.

\begin{figure}
    \centering
    \includegraphics[width=.6\linewidth]{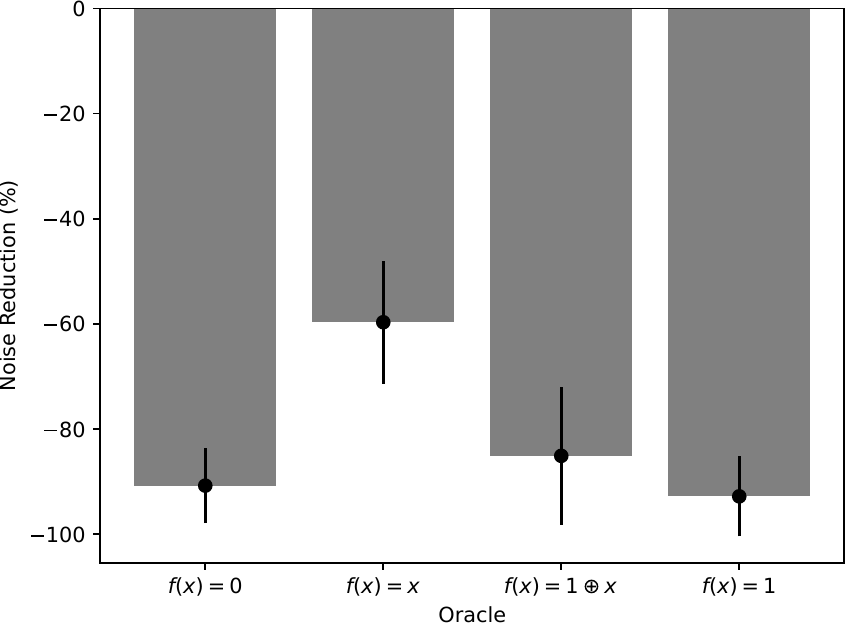}
    \caption{The percentage noise reduction $\left(D_{\rm enc}-D_{\rm bare}\right)/{D_{\rm bare}}$ for all oracles. Error bars denote one standard deviation. We see that there is a statistically significant noise reduction for each oracle.}
    \label{fig:results}
\end{figure}

Results are presented in Table \ref{tab:results}, and the percentage noise reduction is, \begin{equation}
(D_{\rm enc}-D_{\rm bare})/D_{\rm bare}
\end{equation}
is plotted in Figure \ref{fig:results}. We see that there is a statistically significant noise reduction for all four oracles. The worst improvement we observed was for the oracle $f(x)=x$, for which $D_{\rm enc}-D_{\rm bare} = -0.00189 \pm 0.000741$, which is negative with approximately $99 \%$ confidence, and corresponds to a noise reduction of $-(60 \pm 12) \%$. The average percentage noise reduction we found was
\begin{equation}
    (\overline{D}_{\rm enc}-\overline{D}_{\rm bare})/{\overline{D}_{\rm bare}}  \approx -(88.7 \pm 4.6) \%.
\end{equation}

\begin{comment}
    For calculation of standard deviation of bare circuits we use these relations
\begin{align}
    \Delta q_0=\sqrt{q_0-q_0^2} \quad \quad      
    \sigma_{q_0}= \frac{\Delta q_0}{\sqrt{4096}} \\ 
    \Delta q_1=\sqrt{q_1-q_1^2} \quad \quad
    \sigma_{q_1}= \frac{\Delta q_1}{\sqrt{4096}} \\
    \sigma_{bare}=\frac{1}{2}*\sqrt{\sigma_{q_0}^2+\sigma_{q_1}^2}
\end{align}
For calculation of standard deviation of encoded circuits we use relations below where R is defined as no of post selected shots
\begin{align}
    \Delta r_0=\sqrt{r_0-r_0^2} \quad \quad
    \sigma_{r_0}= \frac{\Delta r_0}{\sqrt{R}} \\
    \Delta r_1=\sqrt{r_1-r_1^2} \quad \quad
    \sigma_{r_1}= \frac{\Delta r_1}{\sqrt{R}} \\
    \sigma_{encoded}=\frac{1}{2}*\sqrt{\sigma_{r_0}^2+\sigma_{r_1}^2}
\end{align}
For Standard deviation calculation ($\sigma$) we use the following equation. For 99\% confidence we multiply $\sigma*2.6$ \\ \\ 
\begin{align}
       \sigma=\sqrt{\sigma_{bare}^2+\sigma_{encoded}^2}\\
\end{align}
\end{comment}

%%%%%%%%%%%%%%%%%%%%%%%%%%%%%%%%%%%%%%%%%%%%%%%%%%%%%%%%%%%%%%%%%%%%%%%%%%%%%%%%%%%%%%%%%%%%%%%%%%%%%%%%%%%%

\section{Discussion}
Our results show that a fault-tolerant implementation of the simplest quantum algorithm -- the Deutsch-Josza algorithm -- using the smallest quantum error-detecting code -- the $[[4,2,2]]$ code \cite{Vaidman:1996bg} -- is possible and achieves a reduction in noise for all possible oracles, averaging a nearly $90 \%$ reduction in noise, on a commercially available ion-trap quantum computer.  To our knowledge, this is the first completely fault-tolerant implementation of any quantum algorithm (albeit the simplest one) that demonstrates an unambiguous and substantial reduction in noise.

Our aim in this work was to design and demonstrate the simplest non-trivial example of a completely fault-tolerant quantum computation. The resulting circuits, are, however, so simple that one might question whether or not our results are, in fact, non-trivial. Therefore, we ask, what have we learned from this experiment? 

Prior to carrying out the experiment, we were unsure whether or not we could even achieve a statistically-significant reduction in noise for all oracles, let alone a reduction by a full order of magnitude. At one level, therefore, this work simply illustrates the power of the simplest quantum error-prevention scheme of \cite{Vaidman:1996bg}, and confirms that the basic ideas of fault-tolerant quantum computing not only work, but work very well, not only in theory, but even with present-day technology.

%If, in the future, large-scale fault-tolerant quantum computation is achieved, via, e.g., the surface code and magic state distillation \cite{MSD}, the present work would be of value only as a pedagogical experiment designed to illustrate the basic principles of fault-tolerance. However, given the tremendous investment society has made in realizing noisy, intermediate-scale quantum devices, it seems important to take a more short-term perspective.

More broadly, these results call into question the widely held-belief that error-mitigation is the only practical way to achieve a measurable noise-reduction in the short-term. Error-detection via the $[[4,2,2]]$ code allowed us to achieve noise-reduction exceeding the best error-mitigation schemes, with very low overhead cost -- no ancillae, and an extremely high post-selection ratio of approximately $0.95$. The connectivity requirements of our circuits are also minimal, so they may be executed on various superconducting qubit architectures, such as \cite{Ankaa-2}. 

We also attempted to compare fault-tolerant and non-fault-tolerant preparations of entangled states, following \cite{Vuillot2018}, using the same hardware. We could not demonstrate a statistically significant reduction in noise for all entangled states -- particularly for those entangled states whose preparation required the use of ancillae. To our knowledge, an unambiguous demonstration of noise reduction over all gates, state preparations, and measurements using the $[[4,2,2]]$ code is still an open problem. Our work, however, demonstrates that, even before quantum hardware enables us to achieve this more ambitious goal, small fault-tolerant schemes can achieve, through a judicious choice of logical operators and state preparation circuits, a reduction in noise when executing specific quantum algorithms.

Are the methods used here scalable? In our view, the only truly scalable approach to fault-tolerance in the long-term is to implement Clifford gates via a code such as the surface code, supplemented by magic state distillation \cite{MSD}. Many may also argue that many existing quantum computer architectures, particularly ion-traps, require some new innovations in design for scalability. There is, therefore, always a tradeoff between short-term results and long-term scalability in quantum technology. However, we do believe it is  possible, and would be tremendously interesting, to apply the technique of ``fault-tolerance by error-detection'', supplemented with judicious encoding, perhaps using the $[[n, n-2,2]]$ code and related Bacon-Shor codes, to execute larger instances of the Deutsch-Josza algorithm and other textbook quantum algorithms, aiming for $\sim O(10)$ logical qubits, in the near future. Whether such implementations can be used to achieve a noise reduction as large that obtained in the present work, is an open question.

%Indeed, we also carried out preliminary attempts at running these circuits on Rigetti's 9-superconducting qubit machine, Ankaa-9Q-3, and observed in a similar reduction in noise, which we hope to report on in the future. 

\begin{acknowledgments}
SP thanks Prof. P.S. Satsangi for inspiration and guidance.  We acknowledge the support of MeitY QCAL, Amazon Braket and DST-SERB grant (CRG/2021/009137). We also thank K. Ramya, Tanay Saha, Shubham Sinha and Amolak Kalra for discussions, and the developers of the Qiskit SDK \cite{qiskit2024} which we used when running our quantum circuits.
\end{acknowledgments}

\bibliographystyle{ieeetr}
\bibliography{references}

\appendix

\clearpage

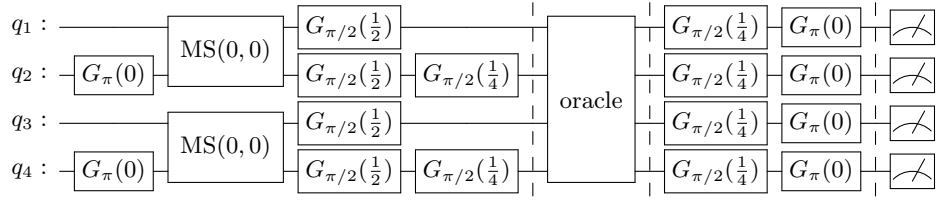
\begin{figure*}
    \centering
    \scalebox{1}{
    \Qcircuit @C=0.6em @R=0.2em @!R { \\
    	 	\nghost{{q}_{1} :  } & \lstick{{q}_{1} :  } & \qw & \multigate{1}{\mathrm{MS}(0,0)}_<<<{} & \gate{G_{\pi/2}(\frac{1}{2})} & \qw \barrier[0em]{3} & \qw & \multigate{3}{\mathrm{oracle}}_<<<{} \barrier[0em]{3} & \qw & \gate{G_{\pi/2}(\frac{1}{4})} & \gate{G_{\pi}(\mathrm{0})} \barrier[0em]{3} & \qw & \meter \\
    	 	\nghost{{q}_{2} :  } & \lstick{{q}_{2} :  } & \gate{G_{\pi}(\mathrm{0})} & \ghost{\mathrm{MS}(0,0)}_<<<{} & \gate{G_{\pi/2}(\frac{1}{2})} & \gate{G_{\pi/2}(\frac{1}{4})} & \qw & \ghost{\mathrm{oracle}}_<<<{} & \qw & \gate{G_{\pi/2}(\frac{1}{4})} & \gate{G_{\pi}(\mathrm{0})} & \qw & \meter \\
    	 	\nghost{{q}_{3} :  } & \lstick{{q}_{3} :  } & \qw & \multigate{1}{\mathrm{MS}(0,0)}_<<<{} & \gate{G_{\pi/2}(\frac{1}{2})} & \qw & \qw & \ghost{\mathrm{oracle}}_<<<{} & \qw & \gate{G_{\pi/2}(\frac{1}{4})} & \gate{G_{\pi}(\mathrm{0})} & \qw & \meter \\
    	 	\nghost{{q}_{4} :  } & \lstick{{q}_{4} :  } & \gate{G_{\pi}(\mathrm{0})} & \ghost{\mathrm{MS}(0,0)}_<<<{} & \gate{G_{\pi/2}(\frac{1}{2})} & \gate{G_{\pi/2}(\frac{1}{4})} & \qw & \ghost{\mathrm{oracle}}_<<<{} & \qw & \gate{G_{\pi/2}(\frac{1}{4})} & \gate{G_{\pi}(\mathrm{0})} & \qw  & \meter \\
    \\ }}
    \caption{Fault-tolerant implementation of the Deutsch-Jozsa Algorithm in IonQ Native Gates.}
    \label{fig:native-encoded-deutsch-jozsa}
\end{figure*}
\section{Hardware Specifications and Native Gates}
\label{app:nativegate}

We ran our circuits on IonQ Aria-1, a 25 qubit universal gate-model ion-trap quantum computer \cite{Aria-1} with all-to-all connectivity. This machine is available via multiple cloud providers including Amazon Braket and Microsoft Azure. 

All the bare and encoded circuits were run with 4096 shots, with error mitigation manually disabled. Encoded circuits were run with native gates \cite{native-gates}, explicitly specified using the option of verbatim compilation available in Amazon Braket \cite{verbatim-compilation} to ensure that post-processing does not destroy fault tolerance. 

The following native gates are available on IonQ machines,
\begin{eqnarray}
    G_\pi(\alpha) &=& \begin{pmatrix} 0 & e^{-2 \pi i \alpha} \\ e^{2 \pi i \alpha} & 0 \end{pmatrix}, \\
    G_{\pi/2}(\alpha)&=& \frac{1}{\sqrt{2}} \begin{pmatrix} 1 & -ie^{-2 \pi i \alpha} \\ -ie^{2 \pi i \alpha} & 1 \end{pmatrix},
\end{eqnarray} 
and the fully-entangling M\o lmer-S\o rensen gate 
\begin{equation}
    \rm{MS}(0, 0) = \frac{1}{\sqrt{2}} \begin{pmatrix} 1 & 0 & 0 & -i \\ 0 & 1 & -i & 0 \\ 0 & -i & 1 & 0 \\ -i & 0 & 0 & 1\end{pmatrix}.
\end{equation}  
Our native gate implementation of the fault-tolerant Deutsch-Josza circuit is shown in Figure \ref{fig:native-encoded-deutsch-jozsa}, and that of the fault-tolerant oracles in Table \ref{tab:fault-tolerant-oracles-native}.

\begin{table}[h]
    \centering
    \begin{tabular}{c c}
    \toprule
        \textbf{\makecell{Binary\\ Function}} & \textbf{\makecell{Encoded Oracle \\ in Native Gates}}  \\ \midrule
        & \\
         { $f(x)=0$}&
    \scalebox{1.0}{
    \Qcircuit @C=0.6em @R=0.2em @!R { 
    	 	\nghost{{q}_{1} :  } & \lstick{{q}_{1} :  } & \qw & \qw\\
    	 	\nghost{{q}_{2} :  } & \lstick{{q}_{2} :  } & \qw & \qw\\
    	 	\nghost{{q}_{3} :  } & \lstick{{q}_{3} :  } & \qw & \qw\\
    	 	\nghost{{q}_{4} :  } & \lstick{{q}_{4} :  } & \qw & \qw\\
    \\ }} \\ 
   & \\
    {$f(x)=x$}&
    \scalebox{1.0}{
    \Qcircuit @C=0.6em @R=0.2em @!R { 
    	 	\nghost{{q}_{1} :  } & \lstick{{q}_{1} :  } & \gate{G_{\pi}(\mathrm{0})} & \gate{G_{\pi}(\frac{1}{8})} & \qw & \qw\\
    	 	\nghost{{q}_{2} :  } & \lstick{{q}_{2} :  } & \gate{G_{\pi}(\mathrm{0})} & \gate{G_{\pi}(\frac{3}{8})} & \qw & \qw\\
    	 	\nghost{{q}_{3} :  } & \lstick{{q}_{3} :  } & \gate{G_{\pi}(\mathrm{0})} & \gate{G_{\pi}(\frac{3}{8})} & \qw & \qw\\
    	 	\nghost{{q}_{4} :  } & \lstick{{q}_{4} :  } & \gate{G_{\pi}(\mathrm{0})} & \gate{G_{\pi}(\frac{1}{8})} & \qw & \qw\\
    \\ }}\\ 
   & \\
    {$f(x)=1\oplus x$}&
    \scalebox{1.0}{
\Qcircuit @C=0.6em @R=0.2em @!R { 
	 	\nghost{{q}_{1} :  } & \lstick{{q}_{1} :  } & \gate{G_{\pi}(\mathrm{0})} & \gate{G_{\pi}(\frac{3}{8})} & \qw & \qw\\
	 	\nghost{{q}_{2} :  } & \lstick{{q}_{2} :  } & \gate{G_{\pi}(\mathrm{0})} & \gate{G_{\pi}(\frac{1}{8})} & \qw & \qw\\
	 	\nghost{{q}_{3} :  } & \lstick{{q}_{3} :  } & \gate{G_{\pi}(\mathrm{0})} & \gate{G_{\pi}(\frac{3}{8})} & \qw & \qw\\
	 	\nghost{{q}_{4} :  } & \lstick{{q}_{4} :  } & \gate{G_{\pi}(\mathrm{0})} & \gate{G_{\pi}(\frac{1}{8})} & \qw & \qw\\
\\ }}\\  
& \\
    {$f(x)=1$}&
    \scalebox{1.0}{
    \Qcircuit @C=0.6em @R=0.2em @!R { 
    	 	\nghost{{q}_{1} :  } & \lstick{{q}_{1} :  } & \gate{G_{\pi}(\frac{1}{2})} & \gate{G_{\pi}(\frac{1}{4})} & \qw & \qw\\
    	 	\nghost{{q}_{2} :  } & \lstick{{q}_{2} :  } & \gate{G_{\pi}(\frac{1}{2})} & \gate{G_{\pi}(\frac{1}{4})} & \qw & \qw\\
    	 	\nghost{{q}_{3} :  } & \lstick{{q}_{3} :  } & \qw & \qw & \qw & \qw\\
    	 	\nghost{{q}_{4} :  } & \lstick{{q}_{4} :  } & \qw & \qw & \qw & \qw\\
    \\ }}\\
    \bottomrule
    \end{tabular}
    \caption{Oracles implemented via native gates for IonQ's Aria-1 machine.}
    \label{tab:fault-tolerant-oracles-native}
\end{table}

\begin{table}
    \centering
    \vspace{1cm}
    \begin{tabular}{l r}
    \toprule
       \multicolumn{2}{c}{\textbf{Calibration Data}} \\ \midrule
       Average 1Q fidelity (\%)  & 99.05 \\ 
       Average 2Q fidelity (\%)  & 98.75 \\  
       Average readout fidelity (\%) & 99.32 \\
        T1 ($\mu$s) & 100000000.000 \\ 
        T2 ($\mu$s) & 1000000.000 \\ 
        1Q gate duration ($\mu$s)&135.000 \\ 
        2Q gate duration ($\mu$s) & 600.000 \\ 
        Readout duration ($\mu$s) & 300.000\\ 
        Active reset duration ($\mu$s) & 20.000 \\
        \bottomrule

    \end{tabular}
    \caption{Calibration data taken at the time of run. ($1Q$ and $2Q$ refer to single-qubit and two-qubit gates. T1 is the time scale for a $\ket{1}$ state to decay toward the ground state $\ket{0}$ and T2 is the time scale for a $\ket{+}$ state to decohere into the completely mixed state.)}
    \label{tab:calibration-data}
\end{table}

The calibration data provided by IonQ, at the time of the run, is shown in Table \ref{tab:calibration-data}.

\section{Entangled States}
While the purpose of this work is not to benchmark any quantum computer, in order to compare our hardware and results to those of \cite{Vuillot2018}, we also prepared entangled states fault-tolerantly within the $[[4,2,2]]$ code. We present the results in this appendix.

\begin{figure}
    \centering
    \includegraphics[width=0.6\linewidth]{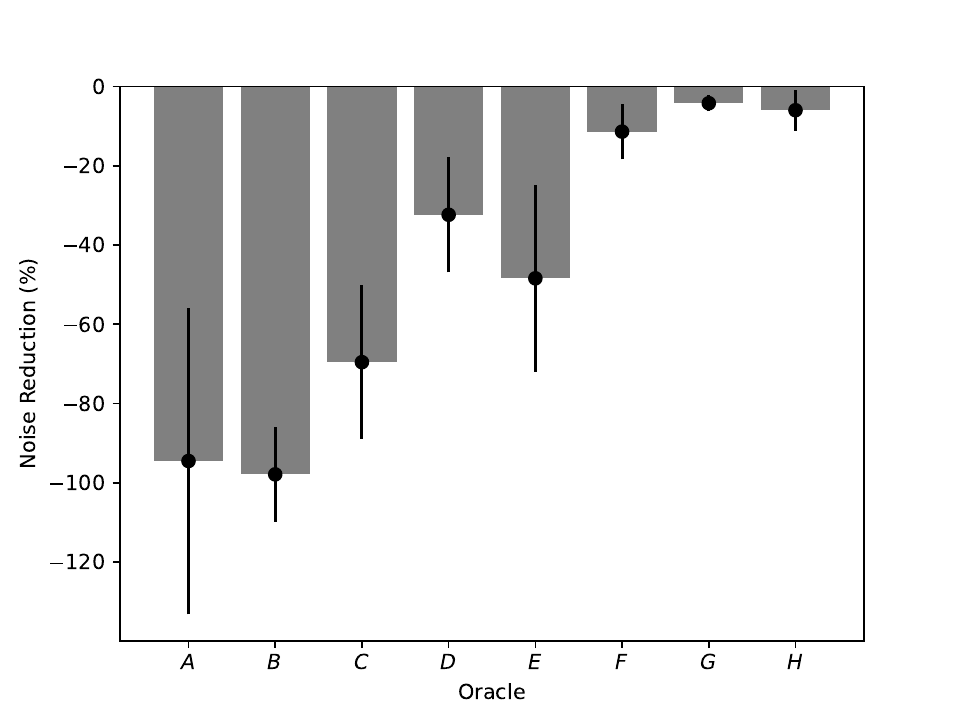}
    \caption{A plot of the percentage noise reduction comparing bare and encoded preparation of 8 entangled-states.}
    \label{fig:results2}
\end{figure}

We followed the same conventions as \cite{Vuillot2018}, and in particular, used the ``conventional'' fault-tolerant dictionary for the $[[4,2,2]]$ code, shown in table \ref{old-dictionary}. The entangled states we prepared are shown in Table \ref{entangled-states-table}. 
Our results are presented in Table \ref{tab:results-entangled}, and the percentage noise reduction shown in Figure \ref{fig:results2}. We see that, while there is a noise reduction in all cases, it is not always statistically significant. 

\begin{table}
\centering
\begin{tabular}{|c|c|c|c|c|}
\hline
\textbf{Logical Gates}&\textbf{Physical Gates}\\ \hline
 $I\otimes I$&$X^{\otimes 4}, Z^{\otimes 4}$ \\
 $X\otimes I$&$X \otimes I \otimes X \otimes I$ \\
$I\otimes X$&$X \otimes X \otimes I \otimes I$  \\
 $Z\otimes I$&$Z \otimes Z \otimes I \otimes I$ \\
$I\otimes Z$&$Z \otimes I \otimes Z \otimes I$  \\
SWAP$_{12} \circ (H\otimes H)$&$H \otimes H \otimes H \otimes H$  \\
 C-Z&$P \otimes ZP \otimes ZP \otimes P$ \\ \midrule
CNOT$_{12}$&SWAP$_{12}$  \\
CNOT$_{21}$&SWAP$_{13}$  \\
SWAP$_{12}$&SWAP$_{23}$  \\
\hline
\end{tabular}
\caption{The map between logical and physical gates for the $$[[4,2,2]]$$ code using the alternative conventions of \cite{gottesman2016quantum, Vuillot2018}.}\label{logicalgates}\label{old-dictionary}
%\end{center}
\end{table} 

\begin{table}[p]
	\renewcommand{\arraystretch}{1.5}
	\centering
	\vspace{0.1cm}
	\begin{tabular}{c c c c}
		\toprule
		Circuit Name & Initial state&Unitary & Final state  \\\midrule
  
		A&$\frac{\ket{00}+\ket{11}}{\sqrt{2}}$ & ${I}\otimes{I}$ & $\frac{\ket{00}+\ket{11}}{\sqrt{2}}$ \\
    
		B&$\frac{\ket{00}+\ket{11}}{\sqrt{2}}$ & ${I}\otimes Z$ & $\frac{\ket{00}-\ket{11}}{\sqrt{2}}$ \\
  
	   C&$\frac{\ket{00}+\ket{11}}{\sqrt{2}}$ & $X\otimes {I}$ & $\frac{\ket{10}+\ket{01}}{\sqrt{2}}$ \\

		D&$\frac{\ket{00}+\ket{11}}{\sqrt{2}}$ &${I}\otimes ZX$ & $\frac{\ket{10}-\ket{01}}{\sqrt{2}}$ \\

		E&$\ket{00}$ & $\mathrm{C}Z\cdot H\otimes H$ & $\frac{\ket{00}+\ket{01}+\ket{10}-\ket{11}}{2}$ \\
  
		F&$\ket{00}$ & $\mathrm{C}Z\cdot {Z}\otimes I\cdot H\otimes H$ & $\frac{\ket{00}-\ket{01}+\ket{10}+\ket{11}}{2}$ \\
  
		G&$\ket{00}$ & $\mathrm{C}Z\cdot I\otimes{Z}\cdot H\otimes H$ & $\frac{\ket{00}+\ket{01}-\ket{10}+\ket{11}}{2}$ \\
  
		H&$\ket{00}$ & ${X}\otimes I\cdot\mathrm{C}Z\cdot H\otimes H\cdot X\otimes{I}$ & $\frac{\ket{00}-\ket{01}-\ket{10}-\ket{11}}{2}$ \\\bottomrule
	\end{tabular}
    \caption{Various entangled states that can be prepared using the $[[4,2,2]]$ code.} \label{entangled-states-table}
\end{table}

\begin{table*}[ht]
	\renewcommand{\arraystretch}{2}
	\centering
	\vspace{0.1cm}
	\begin{tabular}{l  c  c  c  c }
		\toprule
		 \textbf{Circuit}& \textbf{\makecell{Statistical Distance \\ (Bare)}} & \textbf{\makecell{Statistical Distance\\ (Encoded)}}& \textbf{\makecell{Statistical Distance\\ (Encoded-Bare)}} & \textbf{\makecell{Post-Selection \\Ratio}}\\
            \midrule 
		$A$& $0.01367 \pm 0.0056$ & $0.00075 \pm 0.00567$ & $-0.01292  \pm 0.00797$ &0.973\\

		$B$& $0.0471 \pm 0.00576$ & $0.001004 \pm 0.00584$ & $-0.0461  \pm 0.00821$ &0.973\\
	   $C$ &$0.0202 \pm 0.00563$&	$0.00616 \pm 0.00571$	&$-0.0140 \pm 0.00802$&0.969\\

		$D$ &$0.0124 \pm 0.00559$&$0.00841 \pm 0.00567$&$-0.00403 \pm 0.00796$&0.971\\
        
        $E$ &$0.0139 \pm 0.00676$&$0.00717 \pm 0.00699$&$-0.00673 \pm 0.009731$&0.935\\
        $F$ &$0.0109 \pm 0.00676$&$0.00973 \pm 0.00697$&$-0.00125 \pm 0.00971$&0.941\\
        $G$ &$0.0144 \pm 0.00676$&$0.0.0137 \pm 0.00705$&$-0.000611 \pm 0.00977$&0.9204\\
        $H$ &$0.00781 \pm 0.00676$&$0.0.00734\pm 0.00707$&$-0.000469 \pm 0.00978$&0.9143\\
	\bottomrule 
	\end{tabular}
    
    \caption{Results for bare (non-fault tolerant) and encoded (fault-tolerant) implementation of the Entangled States, on IonQ's Aria-1 machine, with 4096 shots in each run.}
    \label{tab:results-entangled}
\end{table*}

\end{document}